\documentclass[preprint]{aastex} 

\begin{document}

\title{ABSOLUTE MAGNITUDE DISTRIBUTION AND LIGHT CURVES OF 
GAMMA-RAY BURST SUPERNOVAE}

\author{Dean Richardson\altaffilmark{1,2,3} \\ 
(richardsond@denison.edu)}  

\altaffiltext{1}{Dept. of Physics and Astronomy, Denison University, Granville, 
OH 43023}  
\altaffiltext{2}{Physics Dept., Marquette University, Milwaukee, WI 53201}
\altaffiltext{3}{Dept. of Physics and Astronomy, University of Oklahoma, Norman,
OK 73019}

\begin{abstract} 
Photometry data were collected from the literature and analyzed for supernovae 
that are  
thought to have a gamma--ray burst association. There are several gamma--ray bursts  
afterglow light curves that appear to have a supernova component. For these 
light curves, the supernova component was extracted and analyzed. A supernova 
light curve model was used to help determine the peak absolute magnitudes as well 
as estimates for the kinetic energy, ejected mass and nickel mass in the explosion.   
The peak absolute magnitudes are, on average, brighter than those of
similar supernovae (stripped--envelope supernovae) that do not have a
gamma--ray burst association, but this can easily be due to a
selection effect.  However, the kinetic energies and ejected masses
were found to be considerably higher, on average, than those of
similar supernovae without a gamma--ray burst association.
\end{abstract} 


\section{Introduction} 

A significant effort was made, over a number of years, to find the origin of 
gamma--ray bursts (GRBs). The discovery of optical afterglows in long--duration 
GRBs allowed for the determination of the GRB's redshift 
\citep[for a review, see][and references therein]{vanparadijs00}.  
This led to the realization that long--duration GRBs are at cosmological distances 
\citep[e.g.,][]{vanparadijs97,metzger97}. 
In this paper, long--duration, cosmological GRBs will be referred to simply as GRBs.  

The light curve of the optical afterglow of GRB~980326 showed a rebrightening at around
two weeks after the burst \citep{bloom99}. This rebrightening, or bump, in the 
light curve was attributed to an underlying supernova (SN). This was the first observational 
indication of an association between GRBs and SNe. 

During April of 1998, an  
unusual gamma--ray burst, GRB~980425, was discovered with its gamma--ray energy several orders of 
magnitude lower than typical GRBs \citep{soffitta98,woosley99}. 
Optical observations of the region led to the discovery of an  
unusual supernova, SN~1998bw, within the error radius of the GRB \citep{galama98}.  
Spectra of SN~1998bw showed it to be a Type~Ic SN
with exceptionally broad lines (Ic--BL). This SN was also unusually bright
for a Type~Ic SN \citep{richardson06}. 
It was thought that GRB~980425 might simply be a normal GRB, but one where the jet was 
viewed off axis \citep[e.g.][]{nakamura99,ioka01,yamazaki03}. However, according to 
late--time radio observations \citep{soderberg04}, even if it was viewed off--axis 
it would still be an unusual GRB. There may be more dim, peculiar GRBs, like GRB~980425,   
than observations would indicate; due to the fact that those at large redshifts are 
difficult to detect.  

The afterglow spectra of GRB~030329 were remarkably similar to  
that of SN~1998bw \citep{stanek03}. The SN associated with this GRB was then 
named SN~2003dh. From this, it was clear  
that at least some long--duration gamma--ray bursts are connected to supernovae.   
Since then, two other SNe~Ic--BL have been spectroscopically confirmed 
to have a GRB connection: SN~2003lw/GRB~031203 \citep{malesani04} and 
SN~2006aj/GRB~060218 \citep{sollerman06,mazzali06b,modjaz06,mirabal06}.  
Not all SNe~Ic--BL have a convincing GRB connection; as is the case for 
SN~1997ef \citep{iwamoto00} and SN~2002ap \citep{mazzali02}. However, this does 
not rule out the possibility of an association with a weak (possibly off--axis) GRB.   
It also seems that there are some long--duration gamma--ray bursts with no  
apparent associated supernova; such as GRBs 060505 and 060614 \citep{fynbo06}. 

Similar to the  bump found in the optical afterglow light curve of GRB~980326 (mentioned above), 
there have been several other GRB afterglow light curves with just such a bump   
\citep[e.g.,][]{reichart99,galama00,sahu00,castrotirado01}. This provides us with 
a number of possible GRB/SNe to consider, even without spectroscopic confirmation.  
These light curves are typically analyzed using a composite model 
\citep[e.g.][]{zeh04, bloom04, stanek05}, discussed below in Section~5.  
Most of the light curves considered here are of this type. 
For alternative explanations for this rebrightening see \cite{fynbo04}
and references therein. 
For a discussion on the possible progenitors of long--duration GRBs, 
see \cite{fryer07}. 
The observed data are discussed in Section~2. Section~3 covers the   
analysis of the light curves while the results are presented in Section~4.  
The conclusions are discussed in Section~5.   

\section{Data} 
The photometry data were collected from the literature. R--band data were 
used for most of the GRBs due to their relatively high redshifts.  
V--band data were used for GRB~980425 ($z = 0.00841$) and GRB~060218 ($z = 0.0331$).  
The photometry references are given in Table~1. 

Distances and extinctions were taken into account in order to convert the 
apparent magnitudes to absolute magnitudes. 
Luminosity distances were calculated from the redshifts using the following 
cosmology:   
H$_0 = 60$ $km\ s^{-1}Mpc^{-1}$, $\Omega_M = 0.3$ and $\Omega_{\Lambda} = 0.7$.  
The distance modulus, $\mu$, for each GRB was calculated and is given in Table~1 
along with the references.  

Foreground Galactic extinction was taken into account according to 
\cite{schlegel98}. These values were taken from NED\footnote{The   
NASA/IPAC Extragalactic Database (NED) is operated by the Jet 
Propulsion Laboratory, California Institute of Technology, under contract with 
the National Aeronautics and Space Administration.}    
and converted from A$_B$ to A$_V$. For host galaxy extinction, the best estimates from 
the literature were used. 
In the case where the host galaxy extinction was estimated 
to be small or negligible (yet still unknown), a value of 0.05 mag was assigned. 
The extinctions are listed in Table~1, along with their references.   

Spectra of SN~1998bw were used to calculate K--corrections. The spectra used were from 
\cite{patat01} and obtained from the SUSPECT 
database\footnote{http://bruford.nhn.ou.edu/$\sim$suspect/}.  

\section{Analysis} 
Three of the light curves used in this study had no significant GRB afterglow 
component at the time of the SN: 
GRB~980425/SN~1998bw, GRB~031203/SN~2003lw and GRB~060218/SN~2006aj.  
One of the light curves (GRB~030329/SN~2003dh) had no noticeable SN bump;  
that is to say that the SN component was masked by a slowly declining afterglow. 
However, spectroscopy revealed a significant SN contribution \citep{stanek03}. 
Most of the light curves had both a SN bump and a significant GRB afterglow component 
at the time of the SN. In order to analyze the SN by itself, the GRB afterglow 
component (and sometimes the host galaxy contribution) needed to be removed.  
If the light curve still included light from the 
host galaxy, and yet there was not sufficient late--time data to determine the 
host galaxy contribution, then that GRB was not included in the study. This 
was the case for GRB~991208.  

Once the host galaxy light had been accounted for, there were still two 
components. The GRB and the SN were treated independent of each other.   
While there is certainly some interaction between the GRB and the SN, 
treating them as independent is a reasonable first order approximation. 
The early--time data, in the resulting light curve, were used to determine the  
contribution of the GRB afterglow. After day one, or day two, this was usually
an unbroken power law. It appears as a straight line on a graph of R vs.\
log(t), as shown in Figure~1. The only exception is GRB~030329, where the 
GRB afterglow contribution was determined solely from the spectra. For all other GRBs, 
there needed to be sufficient early--time data to determine the GRB afterglow 
contribution. If this contribution could not sufficiently be determined from 
the data, then that GRB was not used in this study. This was the case for GRB~010921  
and GRB~000911. 

There were, however, a few exceptions (GRBs 020305, 020410 and 020903).  
The two dimmest SNe in the study are from the light curves of 
GRB~020305 and GRB~020410. If the actual GRB contribution was larger than estimated 
for these SNe (shallower slope), then the SN 
contribution would be smaller making these SNe even dimmer than reported here.  
Even if the GRB contribution was negligible (steeper slope, including a downward break), 
then these two SNe would be brighter than reported here, but would  
still be the two dimmest SNe in the study. The light curve of GRB~020903 was not sufficient 
to get a good estimate of the GRB contribution. However, it was sufficient 
to reasonably determine that the GRB afterglow had a negligible contribution at the 
time of the SN's peak brightness. These three SNe are included because their  
peak absolute magnitudes are still relevant, even if the other information obtained
from the light curve fitting remains highly uncertain (see Table~2). 
In general, the uncertainty in determining the 
GRB afterglow contribution is difficult to quantify and has not been included in 
the peak absolute--magnitude uncertainties given in Table~2. 

\subsection{Model}   
A SN light--curve model was used to help obtain accurate peak absolute magnitudes  
from the resulting light curves. It was also used in determining estimates of the kinetic 
energy, ejected mass and nickel mass for each SN. The model used here is a semi--analytical 
model derived from two already existing models; \cite{arnett82} and \cite{jeffery99}.  
At early times, the Arnett model is used; where the diffusion approximation is valid. 
At late times, the deposition of gamma--rays dominates the light curve, and the Jeffery  
model accounts for this. The basic assumptions are spherical symmetry; homologous expansion; 
radiation pressure dominance at early times; that $^{56}$Ni exists and has a distribution 
that is somewhat peaked toward the center of the ejected matter. Also, optical opacity 
is assumed to be constant at early times and gamma-ray opacity is assumed to be constant 
at late times.     
The combined model is described in detail by \cite{richardson06}.  

The model uses the SN's kinetic energy, ejected mass and nickel mass as parameters. 
After searching a grid of parameter values, a least--squares best fit was used to determine 
the most likely parameter values for each light curve. 
In order to improve the results, the ratio of kinetic energy to ejected mass was 
constrained. SN spectra were used to determine this ratio; however, spectra exist for only 
four of the GRB/SNe in the study. 
These ratios are given in foe M$_{\odot}^{-1}$, where 1 foe $= 10^{51}$ erg.   
The values are 5.0 for SN~1998bw \citep{nakamura01}, 4.8 for SN~2003dh 
\citep{mazzali03}, 4.6 for SN~2003lw \citep{mazzali06a} and 1.0 for 
SN~2006aj \citep{mazzali06b}.  
The average of these four values, 3.8, was used for the other GRB/SNe for which spectra 
were not available.  
Possible consequences of this approximation are discussed below (Section~4.2). 

\section{Results} 

\subsection{Absolute--Magnitude Distribution} 
The peak absolute magnitudes are listed in Table~2. 
The uncertainties shown were obtained by taking into account the uncertainties for each of the 
values in Table~1, as well as the observational uncertainties in the apparent magnitudes.  
Figure~2 shows a histogram of the absolute--magnitude distribution. This distribution has 
an average of M$_{V,peak} = -19.2 \pm 0.2$ and a standard deviation of $\sigma = 0.7$.  
Also shown in this figure is the distribution of stripped--envelope SNe, 
\citep[Fig.~2]{richardson06}. Stripped--envelope SNe (SE SNe) are a combination of SNe Ib, 
Ic and IIb; and those shown here, do not have a GRB association. 
The main difference between these two distributions is that the GRB/SNe are, 
on average, brighter by 0.8 magnitudes. This is likely due to  
a selection effect. In order for most of the GRB/SNe to be detected, they have to be relatively  
bright compared to their GRB afterglow.  
Otherwise, it must be close enough to obtain a spectrum; as was the case with GRB~030329.  
Therefore, any relatively distant SN connected with a GRB that has a bright, or slowly declining,
afterglow will not be detected; especially if the SN is relatively dim. This is why the 
dim GRB/SNe are likely to be undercounted.      
Notice that the two dim GRB/SNe fit well within the SE SN distribution.  

A graph of peak absolute magnitude versus distance modulus is shown in Figure~3.  
The two diagonal dashed lines represent the apparent magnitudes of 16 and 25. 
The horizontal dashed line represents the Type~Ia ridge line, and is shown for comparison. 
The GRB/SNe can be compared to SE~SNe \citep{richardson06}, included in this graph. 
Nearly all of the GRB/SNe were found to have an apparent magnitude dimmer than 16, 
but with a limiting magnitude of  25. This is in contrast to the SE SNe which were 
nearly all found to have an apparent magnitude brighter than 16. Since this is 
related to distance, we see that it is rare to find any nearby GRB/SNe. 
Distant GRB/SNe are discovered because their associated GRBs are extremely bright.  

About half of the GRB/SNe in Figure~3 are near the Type~Ia ridge line. 
The dimmest SN in the study, from GRB~020305, has a very large uncertainty.  
It is still worth including, however, due to the fact 
that it is firmly at the low end of the distribution even with the large uncertainty.  

When compared to other studies, 
the absolute magnitudes given here are, on average, somewhat brighter (after
accounting for different cosmologies). 
For example, see Table~5 from \cite{soderberg06} and Figure~6 
(which is similar to Figure~3 of this paper) from \cite{ferrero06}.  
This difference could possibly be due to the different 
methods for extracting the peak absolute magnitudes.  

\subsection{Light Curves} 
The light curves are presented in Figure~4. Most of these light curves have a  
good range over time, considering the difficulty in separating the SN contribution from 
that of the GRB afterglow (early times) and host galaxy (late times). 

The best estimates for kinetic energy, ejected mass and nickel mass are given in Table~2, 
along with rise times.  
Because spectra exists for only four of these GRB/SNe, E$_k/$M$_{ej}$ values are only 
known for these four. The average value is used for the others. This is a reasonable  
first order estimate; however, a change in this value affects the individual E$_k$ 
and M$_{ej}$ values determined by the model. The trend is that increasing the E$_k/$M$_{ej}$ 
ratio leads to an increase in both the E$_k$ and M$_{ej}$ values obtained in the best 
model fits. This does not significantly affect the M$_{Ni}$ values and  
therefore does not affect M$_{V,peak}$. Also notice, from Table~2, that M$_{V,peak}$ 
and E$_k$ are not correlated. This was the case for SE~SNe as well \citep{richardson06}.    
Note that in the light--curve model, it is assumed that a substantial amount of 
$^{56}$Ni is synthesized in the explosion. 
The decay of this isotope powers the peak of the light curve. Then, $^{56}$Co (which is
a product of the $^{56}$Ni decay) itself decays and powers the tail of the light curve.   

The best fit model light curves are all shown on the same graph in Figure~5, with a 
common time of explosion. This shows, 
as expected, a general trend that the brighter SNe have broader light curves. This, however, 
is not always the case. GRB~030723 is very narrow compared to GRB~970228, yet they have 
similar peak magnitudes.  

\section{Conclusions} 
The average peak absolute magnitude was found to be higher for the GRB/SNe than
for the SE SNe. However, in view of possible selection effects, the difference may 
not be significant.   
There were two GRB/SNe at the dim end of the distribution, but well within the
distribution of SE SNe. 

Most of the SN data analyzed here were taken from GRB afterglow light curves. 
The GRB light was treated as being independent of the SN light and was 
removed so that the resulting light curve could be analyzed as a SN 
light curve. These resulting light curves 
fit quite well with a SN light curve model (Figure~4).   

GRB afterglow light curves that are suspected of having a SN component are usually 
analyzed by a different method. Usually the observed 
data are fit to a composite model, where all of the components are represented;   
GRB, SN and host galaxy \citep[for example]{zeh04}. 
The assumptions made are similar to those made with the separation method
of this paper. The main difference between the two methods   
is in the way the SN is treated. The composite method starts with a
light curve of SN~1998bw (adjusted for the GRBs redshift). 
Two of the five free parameters 
in the composite model are then used in the overall fit to describe the SN; 
one for the brightness and one for the width of the SN contribution.  
While the separation method uses SN~1998bw for K--corrections, a general SE SN model 
is then fit to the resulting light curve.  
The two methods are similar; however, the separation method used here allows 
for closer analysis of the the SN. Reasonable estimates of the kinetic energy,   
ejected mass, nickel mass and rise times are found. However, the values of kinetic 
energy and ejected mass from other studies \citep{nomoto06} tend to be larger by 
a factor of approximately two.  

The coincidence of the GRB date and the derived date of the SN explosion is another point 
of interest. In all but two cases the difference between the two dates was less than a week. 
For GRB~050525A, the SN explosion date is estimated to have occurred about 10 days before the 
GRB date, but by looking at the model fit and the observed data in Figure~4, we see that 
the model was not able to simultaneously reproduce the narrow peak and the bright, late--time  
data point. It appears that a more accurate explosion date for the SN would bring it  
closer to the GRB date. For GRB~011121, the SN explosion date is estimated to    
have occurred about 9 days before the GRB date. Other studies have given similar 
results \citep{bloom02}, but the lack of pre--peak data for this SN makes the SN 
explosion date difficult to accurately pin down. Thus there is no clear evidence for real
differences between the times of the GRBs and the SNe.   

The nickel masses determined here range from 0.05 to 0.99
M$_{\odot}$. These are similar to the values found for SE SNe with no
GRB association \citep{richardson06}.  However, the kinetic energy
values determined here range from about one to 31 foe and the ejected
mass values range from about one to six M$_{\odot}$.  These values
are, on average, considerably higher than those found for SE SNe with
no GRB association.

{}

\begin{deluxetable}{lcccccc}  
\tabletypesize{\scriptsize} 
\tablecaption{Data Used In Determining Absolute Magnitudes \label{table1}} 
\tablewidth{0pt} 
\tablehead{ 
\colhead{GRB/} &
\colhead{Photometry} & 
\colhead{$\mu^a$} &
\colhead{Ref.} & 
\colhead{A$_{R}(Galactic)^b$} &
\colhead{A$_{V}(host)$} & 
\colhead{Ref.} 
\\
\colhead{XRF} &
\colhead{Ref.} & 
\colhead{(mag)} & &
\colhead{(mag)} & 
\colhead{(mag)} &  
} 
\startdata 
970228  & 1 &  43.463 $\pm$ 0.003 & 2 & 0.543 $\pm$ 0.087 &  0.15 $\pm$ 0.15 & 3  \\ 
980425  & 4,5,6 &  33.13 $\pm$ 0.26 & 7 & 0.194 $\pm$ 0.031$^c$ &  0.05 $\pm$ 0.05 & 8 \\ 
990712  & 9,10 &  42.224 $\pm$ 0.005 & 11 & 0.090 $\pm$ 0.014  & 0.15 $\pm$ 0.1 & 12 \\ 
011121  & 13,14 &  41.764 $\pm$ 0.006 & 14 & 1.325 $\pm$ 0.212 &  0.05 $\pm$ 0.05 & 14  \\ 
020305  & 15 &  40.29 $\pm$ 1.09 & 15 & 0.142 $\pm$ 0.023 &  0.05 $\pm$ 0.05 & 15 \\ 
020405  & 16 &  43.463 $\pm$ 0.016 & 16 & 0.146 $\pm$ 0.023  & 0.15 $\pm$ 0.15 & 17 \\ 
020410  & 18 &  42.60 $\pm$ 0.43 & 18 & 0.398 $\pm$ 0.064 &  0.05 $\pm$ 0.05 & 18  \\ 
020903  & 19 &  40.844 $\pm$ 0.003 & 19 & 0.0935 $\pm$ 0.0150  & 0.26 $\pm$ 0.26 & 19 \\ 
030329  & 20 &  39.88  $\pm$ 0.01 & 21 & 0.0675 $\pm$ 0.0108  & 0.39 $\pm$ 0.15 & 17 \\ 
030723  & 22 &  43.08 $\pm$ 0.72 & 23 & 0.0886 $\pm$ 0.0142 &  0.23 $\pm$ 0.23 & 22 \\ 
031203  & 24 &  38.775 $\pm$ 0.003 & 25 & 2.772 $\pm$ 0.444 &  0.05 $\pm$ 0.05 & 26 \\ 
041006  & 27 &  43.55 $\pm$ 0.03 & 27 & 0.0607 $\pm$ 0.0097 &  0.11 $\pm$ 0.11 & 17 \\ 
050525A & 28 &  43.10 $\pm$ 0.07 & 29 & 0.2546 $\pm$ 0.0407 &  0.12 $\pm$ 0.06 & 30 \\ 
060218  & 31 &  36.15 $\pm$ 0.05 & 31 & 0.471 $\pm$ 0.075$^c$ &  0.13 $\pm$ 0.13 & 31 \\ 
\enddata 
\tablenotetext{a} {$Luminosity\ Distance$ ($H_0=60$, $\Omega_M=0.3$, $\Omega_{\Lambda}=0.7$)} 
\tablenotetext{b} {\cite{schlegel98}}  
\tablenotetext{c} {A$_V$(Galactic) was used.}  
\tablerefs{
(1) \cite{galama00}, (2) \cite{bloom01}, (3) \cite{castander99}, 
(4) \cite{galama98}, (5) \cite{mckenzie99}, (6) \cite{sollerman00},     
(7) \cite{kay98}, 
(8) \cite{nakamura01}, (9) \cite{sahu00}, (10) \cite{fruchter00}, (11) NED,  
(12) \cite{christensen04}, 
(13) \cite{bloom02}, (14) \cite{garnavich03}, (15) \cite{gorosabel05},     
(16) \cite{masetti03}, (17) \cite{kann06},  
(18) \cite{levan05}, (19) \cite{bersier06}, (20) \cite{matheson03},   
(21) \cite{greiner03}, (22) \cite{fynbo04}, (23) \cite{tominaga04}, (24) \cite{malesani04},   
(25) \cite{prochaska04}, (26) \cite{mazzali06a}, (27) \cite{stanek05}, (28) \cite{dellavalle06},    
(29) \cite{foley05}, (30) \cite{blustin06}, (31) \cite{sollerman06}  
} 
\end{deluxetable} 

\begin{deluxetable}{lcccccccc}  
\tabletypesize{\scriptsize} 
\tablecaption{Results \label{table2}} 
\tablewidth{0pt} 
\tablehead{ 
\colhead{GRB/} &
\colhead{M$_{V,peak}$} &
\colhead{E$_k$} &
\colhead{M$_{ej}$} &
\colhead{M$_{Ni}$} &
\colhead{t$_{rise}$} &
\colhead{N}  \\
\colhead{XRF} &
\colhead{(mag)} &
\colhead{(foe)} &
\colhead{(M$_{\odot}$)} &
\colhead{(M$_{\odot}$)} &
\colhead{(days)} &
} 
\startdata 
970228  & $-18.9 \pm 0.3$ & 23.2 & 6.11 & 0.54 & 24 & 4  \\ 
980425  & $-19.4 \pm 0.3$ & 31.0 & 6.22 & 0.78 & 23 & 104  \\ 
990712  & $-18.9 \pm 0.1$ & 5.32 & 1.40 & 0.20 & 15 & 5  \\ 
011121  & $-19.6 \pm 0.3$ & 14.2 & 3.73 & 0.73 & 21 & 9  \\ 
020305  & $-17.6 \pm 1.1$ & 3.61$^a$ & 0.95$^a$ & 0.05 & 14$^a$ & 7  \\ 
020405  & $-19.8 \pm 0.2$ & 11.1 & 2.92 & 0.72 & 19 & 5  \\ 
020410  & $-18.0 \pm 0.4$ & 6.42$^a$ & 1.69$^a$ & 0.10 & 16$^a$ & 3  \\ 
020903  & $-19.5 \pm 0.3$ & 11.9$^a$ & 3.13$^a$ & 0.60 & 20$^a$ & 11  \\ 
030329  & $-19.5 \pm 0.2$ & 17.4 & 3.63 & 0.62 & 20 & 17  \\ 
030723  & $-19.1 \pm 0.8$ & 3.23 & 0.85 & 0.19 & 13 & 9   \\ 
031203  & $-19.9 \pm 0.5$ & 21.0 & 4.56 & 0.99 & 21 & 10   \\ 
041006  & $-19.9 \pm 0.2$ & 14.5 & 3.81 & 0.94 & 21 & 5   \\ 
050525A & $-18.9 \pm 0.2$ & 15.7 & 4.14 & 0.40 & 21 & 7   \\ 
060218  & $-19.2 \pm 0.2$ & 0.89 & 0.89 & 0.30 & 16 & 42   \\ 
\enddata 
\tablenotetext{a} {These values are highly uncertain due to the difficulty 
in determining the contribution of the GRB afterglow.}  
\end{deluxetable} 

\begin{figure}
\plotone{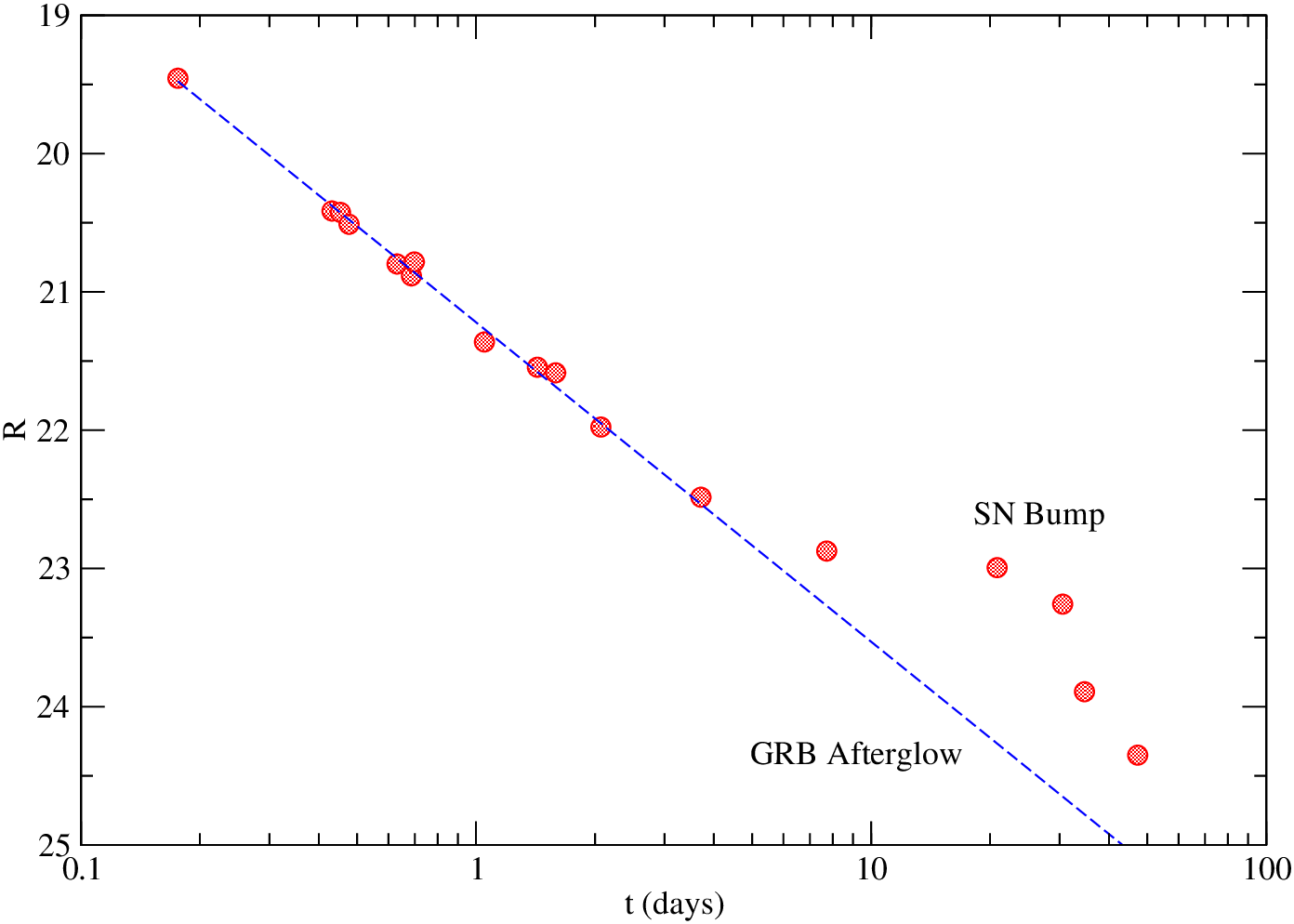}
\caption{\label{fig1}
The observed R--band light curve of GRB~990712 after the host galaxy light had been subtracted. 
The diagonal dashed line represents the contribution due to the GRB afterglow. In this
particular case, the afterglow light can be described by the equation: 
$R_{AG} = 2.31log(t) + 21.22$. 
}
\end{figure} 

\begin{figure}
\plotone{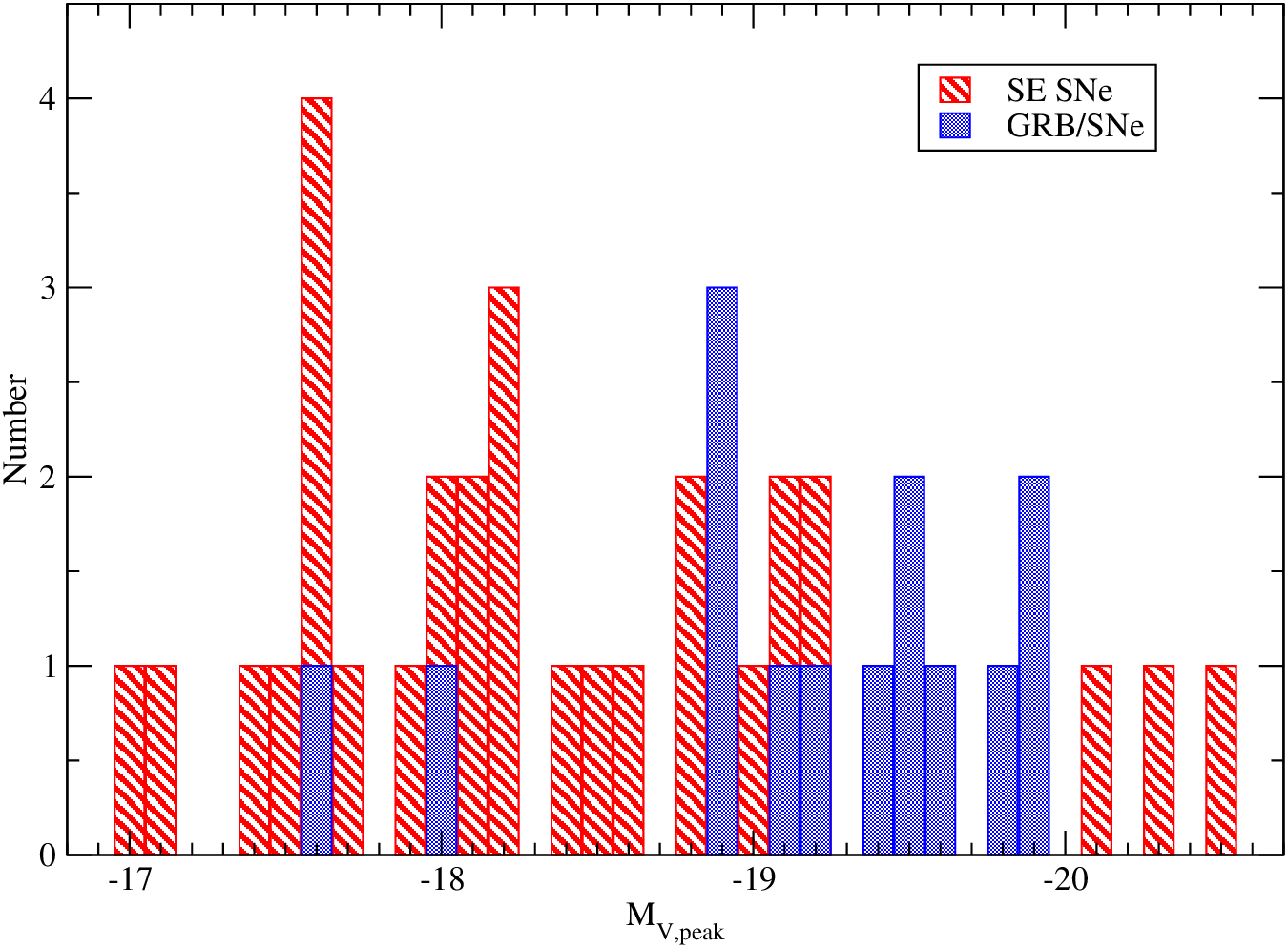}
\caption{\label{fig2}
The peak absolute--magnitude distribution for GRB/SNe (solid bars) is shown with an average 
value of $M_{V,peak} = -19.2 \pm 0.2$ and a standard deviation of $\sigma = 0.7$.  
SE SNe are shown for comparison (striped bars). 
}
\end{figure} 

\begin{figure}
\plotone{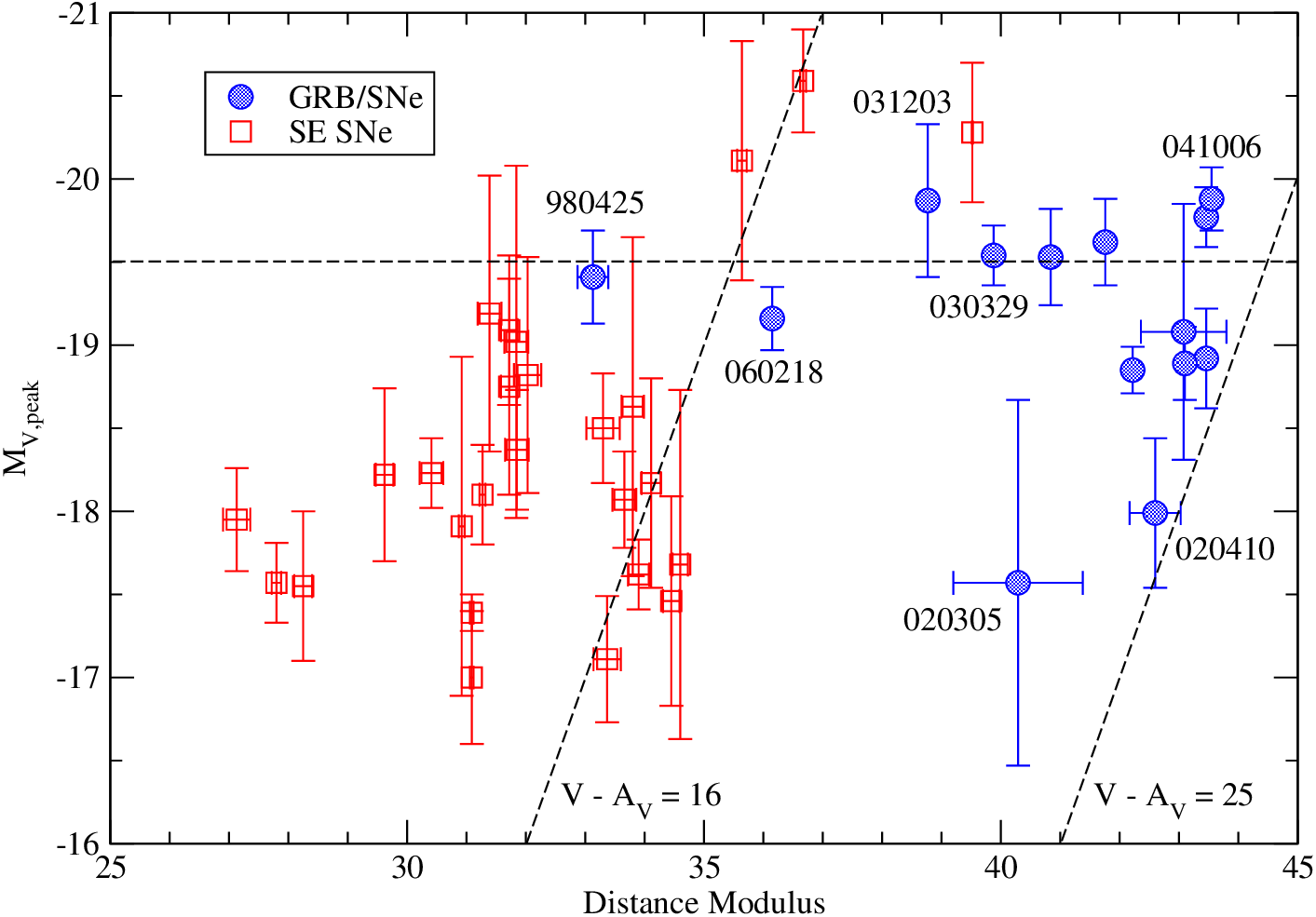}
\caption{\label{fig3}
Peak absolute--magnitude is plotted here versus distance modulus. 
The diagonal dashed lines 
are lines of constant apparent magnitude (16 and 25 mag.). 
The dashed horizontal line at $M_V = -19.5$,  
representing the SN~Ia ridge line, is shown for comparison. 
The filled circles are GRB/SNe and the open squares are SE SNe.  
The associated GRB names are used to label a few key GRB/SNe.  
}
\end{figure} 

\begin{figure}
\plotone{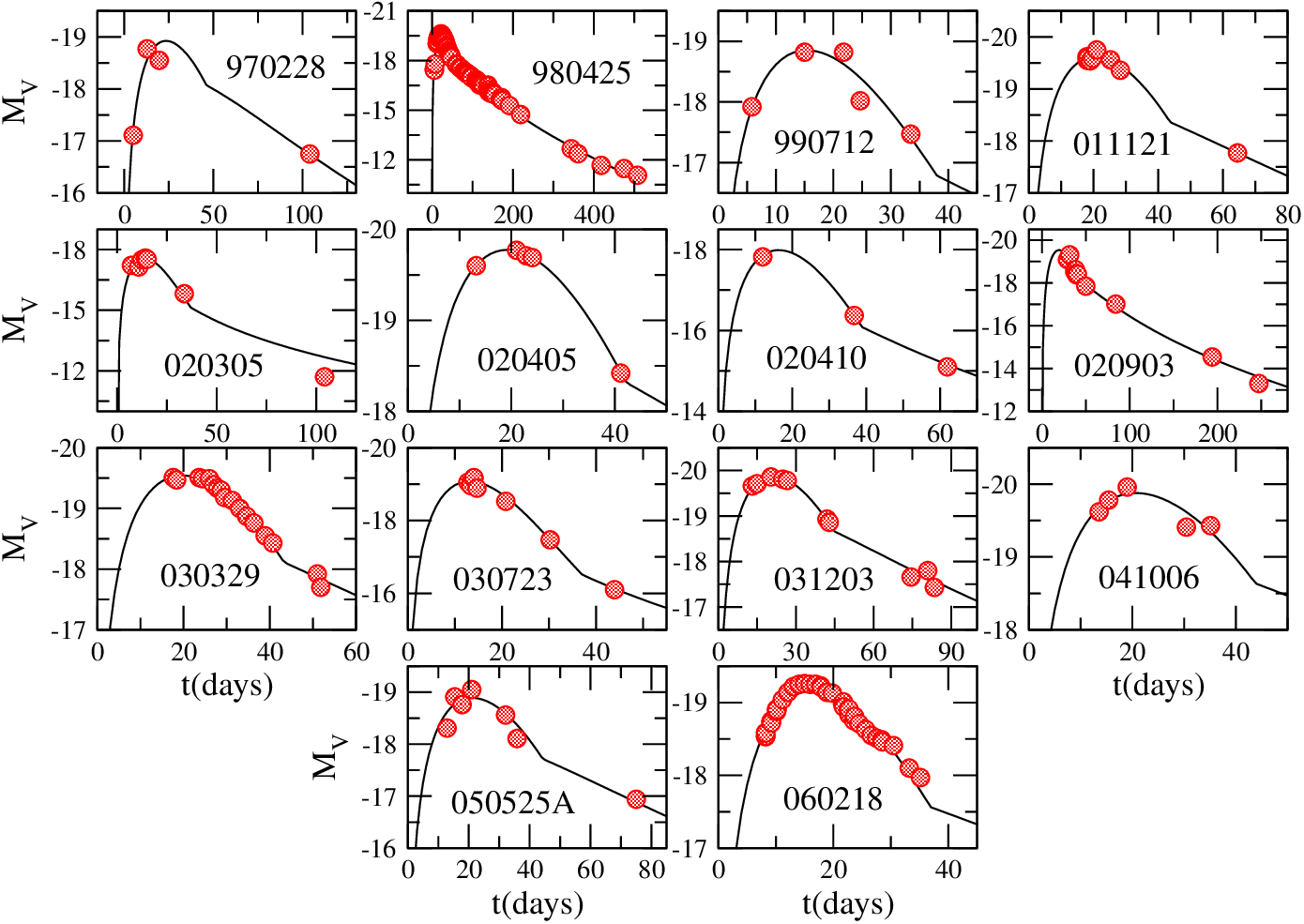}
\caption{\label{fig4}
All of the GRB/SN light curves in the study are plotted here with the best model fits. 
}
\end{figure} 

\begin{figure}
\plotone{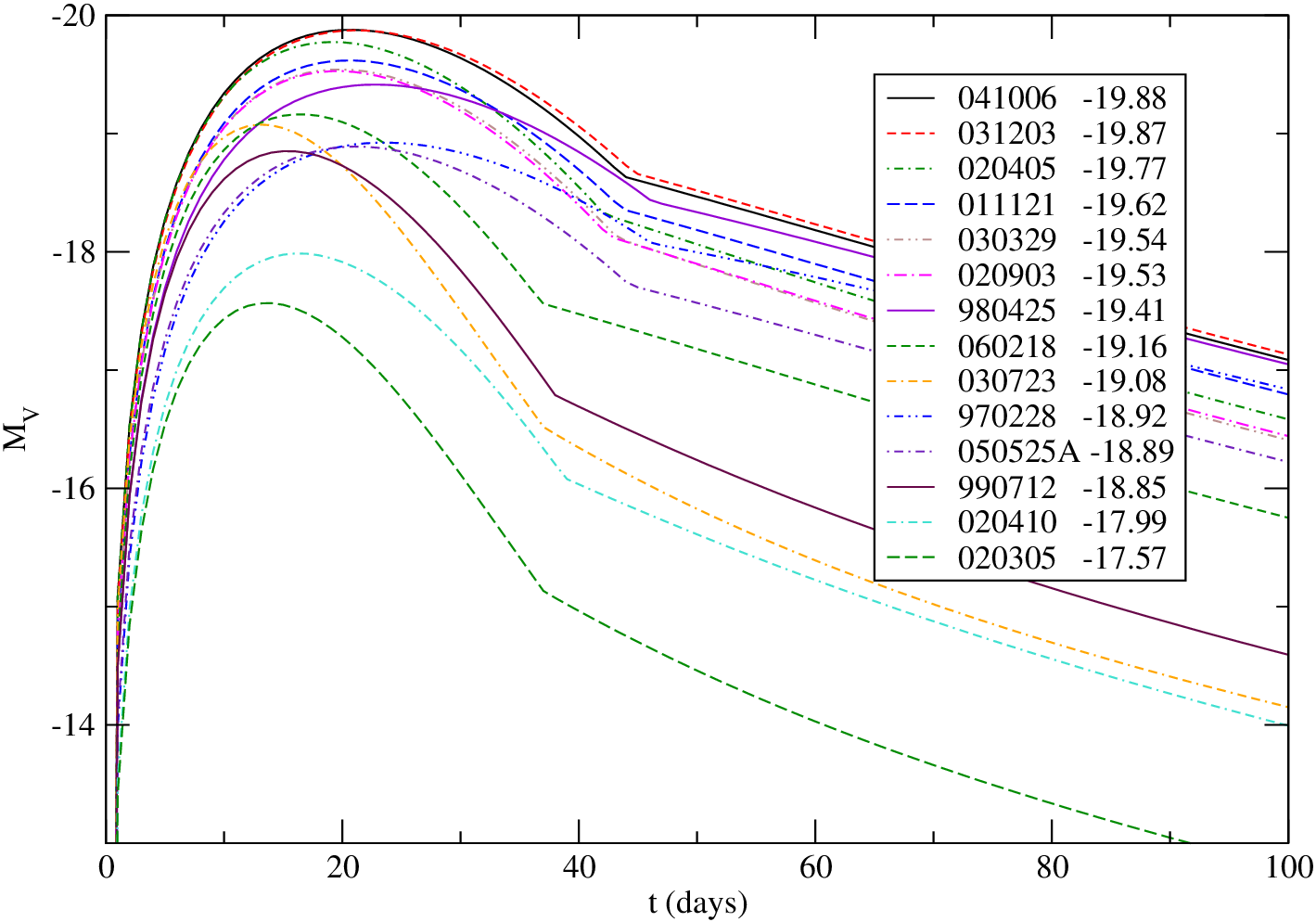}
\caption{\label{fig5}
All of the GRB/SN model light curves in the study are plotted here. The peak absolute 
magnitudes are given for each.  
}
\end{figure}

\end{document}